\documentclass[preprint,5p]{elsarticle}
\usepackage{braket}
\usepackage{cancel}
\usepackage{titlesec}
\usepackage[utf8]{inputenc}
\usepackage{dcolumn}
\usepackage{multirow}
\usepackage{here}
\usepackage{color}

\begin{document}
\biboptions{sort&compress}
\title{Phonon dynamics of a bulk WSe$_2$ crystal excited by \textcolor{black}{ultrashort near-}infrared pulses}

\author[{1,2}]{Itsuki Kasai}
\author[{1,2}]{Itsuki Takagi}
\author[{1,2}]{Kazutaka G. Nakamura}
\ead{nakamura@msl.titech.ac.jp}

\affiliation[1]{Materials and Structures Laboratory, Institute of Science Tokyo,4259 Nagatsuta, Yokohama 226-8501, Japan}
\affiliation[2]{Department of Materials Science and Engineering, Institute of Science Tokyo, 4259 Nagatsuta, Yokohama 226-8501, Japan}
\cortext[cor1]{Corresponding author}

\begin{abstract}
Pump-probe reflectivity measurements have been performed on a single crystal of tungsten diselenide (WSe$_2$) using ultrashort near infrared pulses.
The behavior is well reproduced in simulations superimposing three oscillations (7.45, 7.49 and 7.7 THz) with different phases.
The Fourier transform spectrum features small peaks at 4.0 and 11.5 THz along with intense peaks at around 7.5 THz.
\end{abstract}


\maketitle

\section{Introduction}
Ultrafast dynamics of optical phonons are widely studied via coherent phonons, which are generated by irradiation of ultrashort optical pulses, using a transient reflection and transmission measurements.\cite{Forst2008} 
The change in reflected light intensity ($\Delta$R$/$R) oscillates as a function of the pump--probe delay, with a frequency corresponding to optical phonon frequencies.
Frequency, relaxation and decoherence processes in various materials (for examples, semiconductors, superconductors and topological insulators) are typically investigated by analyzing the transient reflection behavior.

Transition metal dichalcogenides (TMDCs) such as MoSe$_2$ and WSe$_2$ have recently attracted much attention in
optoelectronic device materials due to their unique optical and electrical properties including high carrier mobility \cite{Qiu2021} and superconductivity\cite{Ahmed2017}.
TMDCs have a two-dimensional layered structure held together by van der Waals interaction, with band gap and phonon frequencies that vary depending on the number and orientation of the layers \cite{Zhao2013, Pan2022}.
Calculations and spectroscopy have shown that the electron--phonon interactions is an important factor for dominating these properties.\cite{Chen2021, Bawden2016} 
Optical phonons have been extensively studied using Raman spectroscopy.
The bulk WSe$_2$ crystal has point group symmetry of $D_{6h}^4$ and exhibits Raman active vibrational modes denoted as A$_{1g}$, E$_{1g}$, and 2E$_{2g}$ at the $\Gamma$ point.
The most intense peak, reported at 250 cm$^{-1}$, corresponds to overlapped E$_{2g}$- and A$_{1g}$-mode phonons.
The number of observed peaks and their frequencies depend on excitation wavelengths and the number of layers.\cite{Blaga2024}
Coherent phonons in WSe$_2$ have also been investigated by transient reflection or transmission measurements using tens-of-femtosecond near-infrared pulses \cite{Jeong2016, Jeong2020}.
A coherent oscillation with a frequency of 7.5 THz (corresponding to 250 cm$^{-1}$) has been observed in the transient transmission light intensity, with a reported lifetime of approximately 7.0 ps at room temperature.
Additionally, the Fourier-transformed spectrum of the transient oscillation features a small peak at 4.0 THz along with the intense peak at 7.5 THz.
However, the observed coherent phonon oscillations are quite limited compared with Raman measurements.

In this paper, we investigate coherent phonons in a bulk crystal of WSe$_2$ using transient reflection measurements with \textcolor{black}{ultrashort near-infrared} pulses.
The use of shorter pulses allows for the excitation and observation of higher-frequency phonons.
In addition to the intense 7.5-THz phonons, a peak at 11.5 THz is also found in Fourier-transformed spectrum of the transition reflection signals.
Furthermore, the transient reflection measurement revealed that its intensity increases to approximately 1 ps after the pump pulse irradiation before decaying.
We propose that this increase may be explained by the superposition of three damped oscillations with different phases. 

\section{Experimental}

The transient reflection signals were investigated using a pump--probe technique with near infrared femtosecond pulses.
The output from the Ti:sapphire oscillator (FEMTO-LASER Rainbow) was directed through a pair of chirped mirrors to compensate for group-velocity dispersion of the optical path.
\textcolor{black}{The pulse width was measured to be $\sim$ 20 fs in this experiments.}
The pulse was the divided into two pulses (3:1 ratio) by a beam splitter.
The pump pulse was directed to a scanning delay unit (APE scanDelay 50) to control the delay between pump and probe pulses.
 Both pulses were focused on the sample using a parabolic mirror, with the pump power set at approximately 90 mW.
 The reflected probe pulse was directed to a polarizing beam splitter, and the s- and p-polarized light was detected by a pair of balanced photo detectors.
 The differential signals were amplified using a current amplifier (Stanford Research: SR570) with electrical band pass filter between 3 kHz and 10 kHz.
 The signals were stored and averaged by a digital oscilloscope (Iwatsu: DS5634A).
 Further details of the experimental setup are described elsewhere \cite{Nakamura2016}.
  
 The sample was a commercially available WSe$_2$ crystal (produced by HG Graphene, obtained from EM Japan Co.).
 Raman spectra were measured using a Laser Raman spectrometer (JASCO NRS-4100) with a 50$\times$ objective lens and an excitation wavelength of 784.7 nm (continuous wave laser).
 
 \section{Results}

\begin{figure}[t]
    \centering
    \includegraphics[width=8cm]{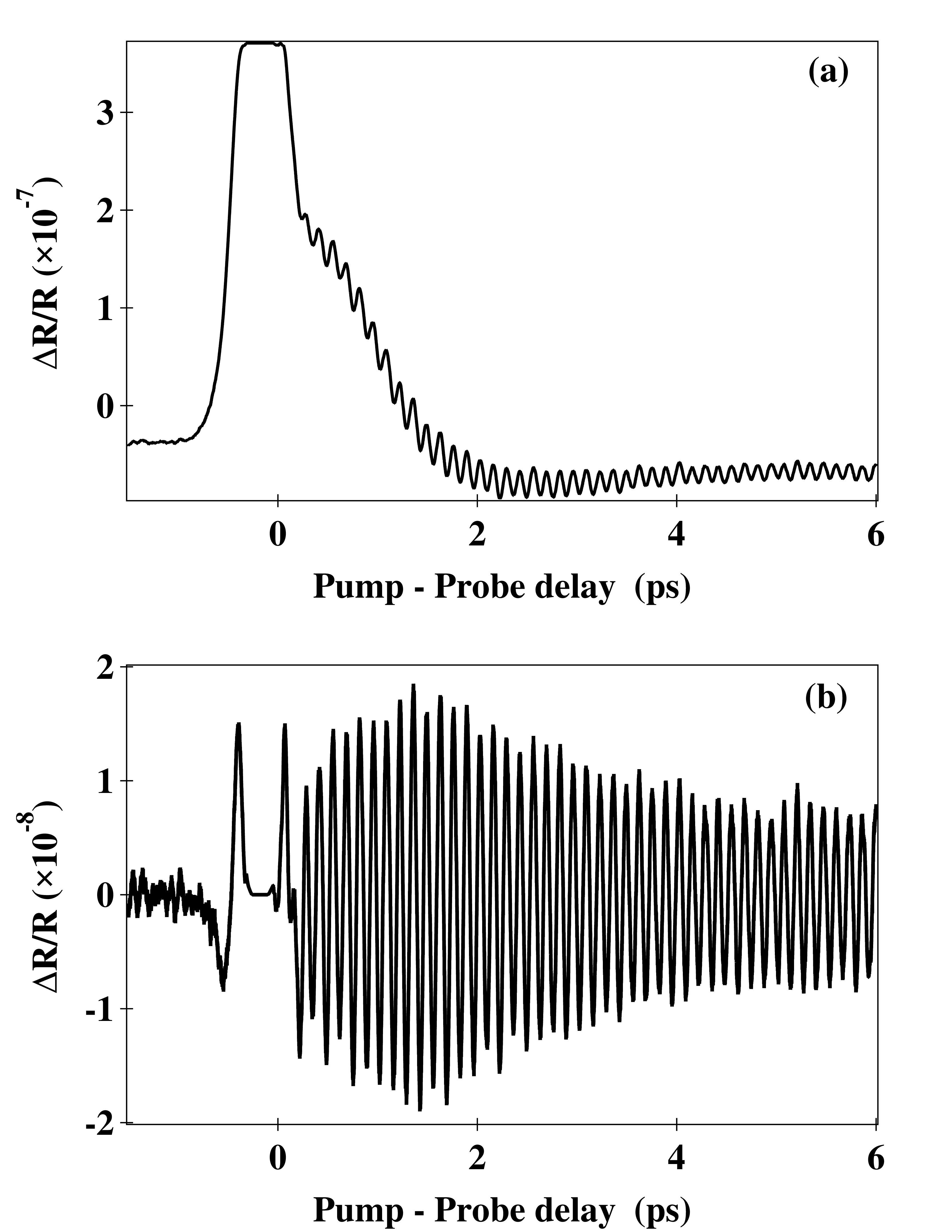}
        \vspace{1cm}
    \caption{Transient reflectivity measurement of WSe$_2$; (a) transient reflection as a function of the delay between pump 1 and probe pulses; (b) oscillational part of (a), which is obtained by subtracting the baseline.}
    \label{reflection}
\end{figure}

Figure \ref{reflection}(a) shows the transient reflectivity of WSe$_2$ as a function of the pump-probe delay ($\tau$).
The reflectivity exhibits a strong positive response at $\tau =0$, with a weak oscillation superimposed on the intense response.
Figure \ref{reflection}(b) shows the oscillational components in the transient reflectivity obtained by subtracting the baseline.
The oscillation has a period of approximately 130 fs, with the oscillation amplitude gradually increasing between 0 and approximately 1 ps before exponentially decreasing. 
 
\begin{figure}[t]
    \centering
    \includegraphics[width=8cm]{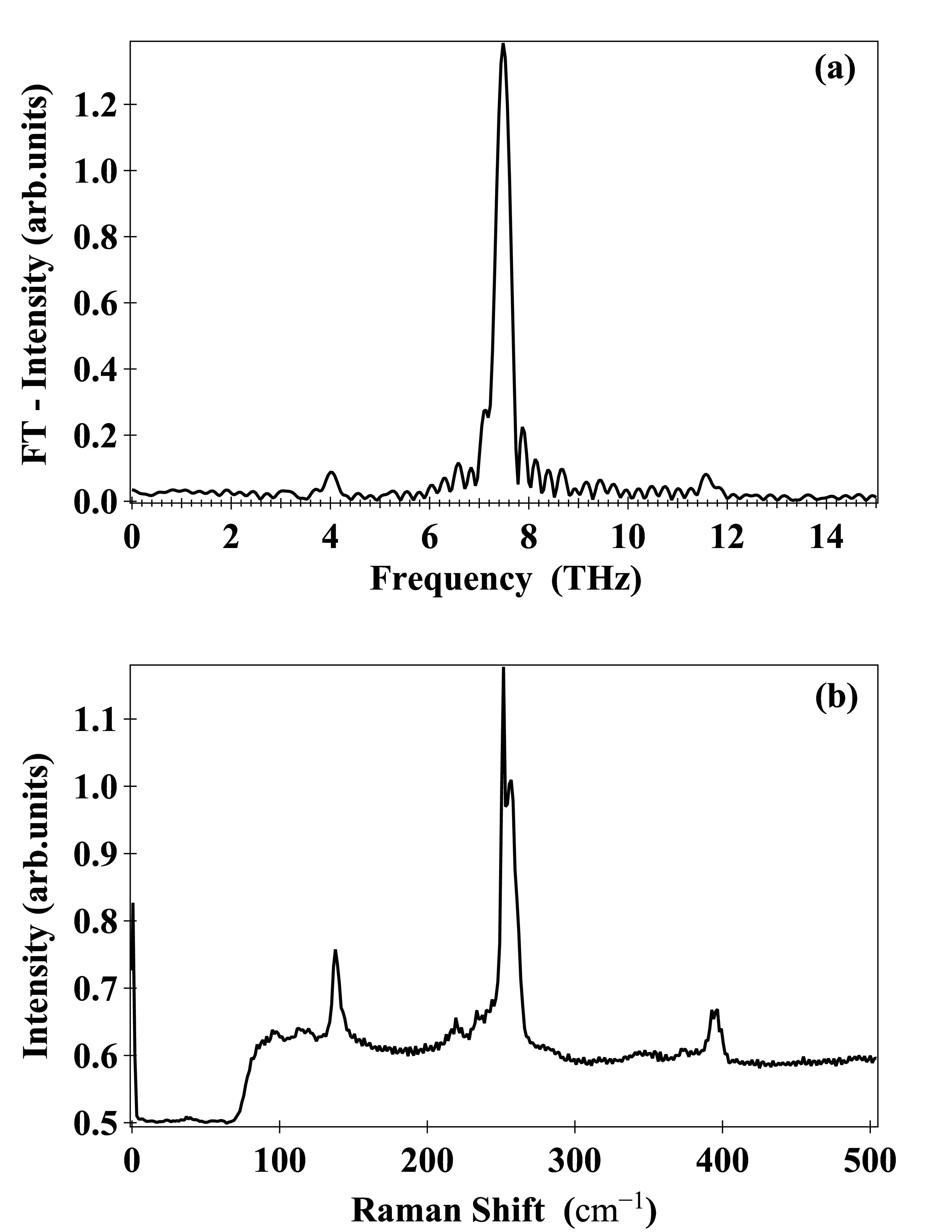}
        \vspace{1cm}
    \caption{(a) Fourier transform spectrum of the transient reflectivity [Fig. 1 (b)]; (b) Raman spectrum obtained with excitation light of 785 nm.}
    \label{spectral}
\end{figure}

The Fourier transform of the transient reflectivity in a delay range between 0.2 and 4.0 ps reveals an intense peak at \textcolor{black}{7.48 THz} and weak peaks at 4.00 and 11.56 THz [Fig. \ref{spectral}(a)].
The frequency resolution is 0.043 THz.
The intensities of the 4.0 and 11.5 THz oscillations are too weak to be identified in the time domain signals [Fig. 1 (b)].
The peaks at 4.00, 7.48, and 11.56 THz are hereafter referred to as the low-frequency, main, and high-frequency peaks, respectively.

Figure \ref{spectral}(b) shows a Raman spectrum obtained using a conventional Raman spectrometer.
The spectrum consists of two intense peaks at 251.67 and 256.84 cm$^{-1}$ (corresponding to 7.545 and 7.700 THz) accompanied by weak peaks at 137.81 and 393.99 cm$^{-1}$  (corresponding to frequencies 4.132 and 11.812 THz).
The Raman spectrum supports that the weak peaks in the FT spectrum are not noise but due to phonon oscillations.

\begin{figure}[t]
    \centering
    \includegraphics[width=9cm]{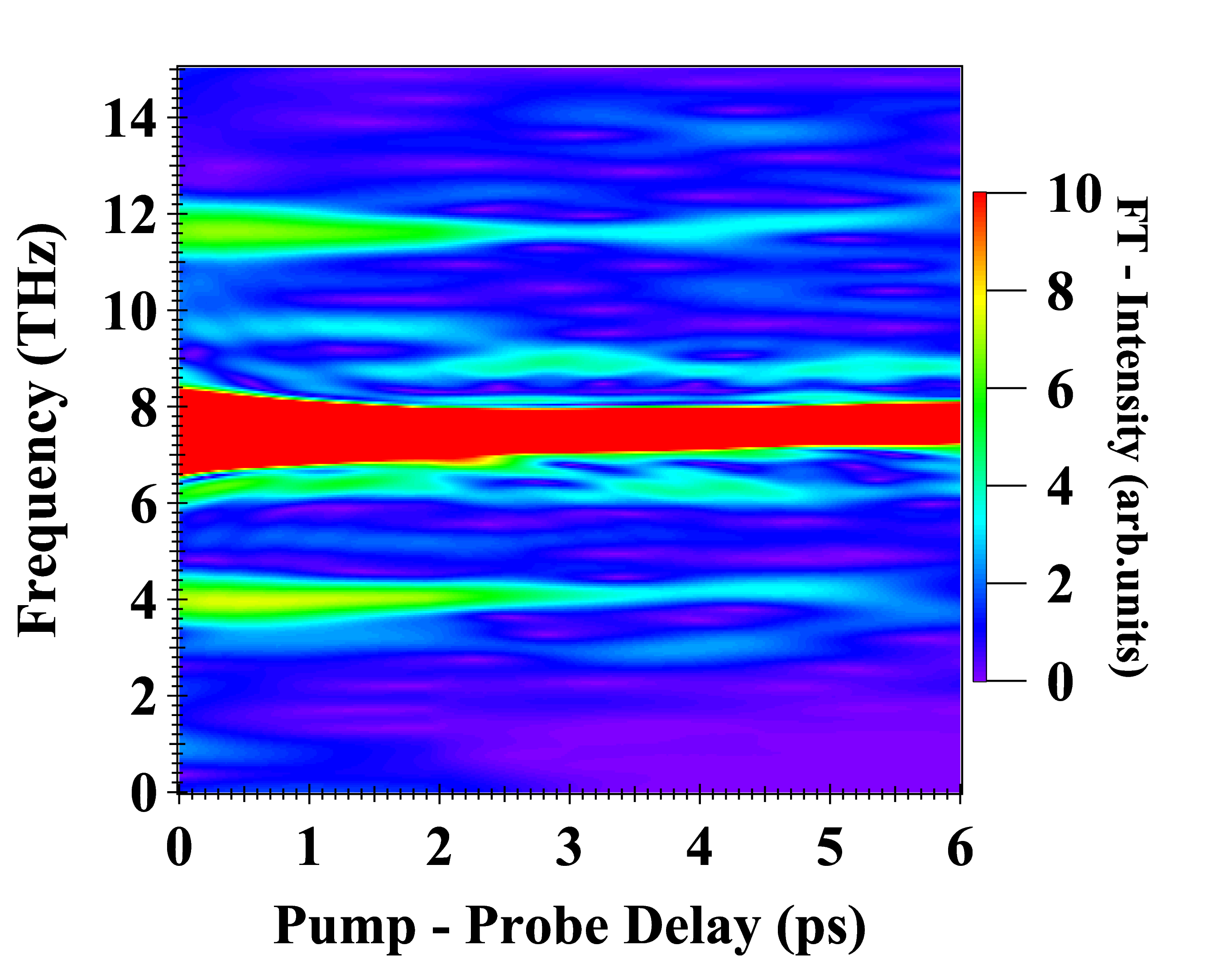}
        \vspace{1cm}
    \caption{Short-time Fourier transform (STFT) spectrum of the transient reflectivity [Fig. 1(b)].}
    \label{STFT}
\end{figure}

To investigate the phonon dynamics in detail, we performed a short-time Fourier-transform (STFT) analysis of the transient reflectivity signals.
The segment size was set to 21410 points (corresponding to approximately 3.8 ps) out of a total size of 79679 points with a step of 10 points.
As shown in Fig. \ref{STFT}(a), the main peak persists over the entire range, whereas the low- and high-frequency peaks are weak and hidden by noise at approximately 4 ps.

\section{Discussion}

The transient reflection signal shows an intense oscillation with a frequency of 7.48 THz, corresponding to the Raman peak at 251 cm$^{-1}$ [Fig. \ref{spectral}(b)].
This oscillation, previously reported by Jeong et al. \cite{Jeong2016}, has been assigned to the A$_{1g}$ phonon.
In earlier Raman measurements, the peak at 250 cm$^{-1}$ (7.5 THz) was attributed to overlapped E$_{2g}$ and A$_{1g}$ modes, whereas the peak at 257 cm$^{-1}$ (7.7 THz) was assigned to two-phonon scattering due to a van Hove singularity (vHs) \cite{Terrones2014, Blaga2024}.
Although the E$_{2g}$ and A$_{1g}$ modes are not degenerated, their energy difference has been calculated to be very small (approximately 0.5 cm$^{-1}$).
Therefore, the intense 7.48 THz oscillation might be attributed to the overlapped E$_{2g}$ and A$_{1g}$ modes, with the 7.7-THz oscillation, potentially contributing to the coherent oscillation.

The coherent oscillation in the transient reflection exhibits a significant characteristic: an intensity rise near zero delay ($\tau=0$).
Typically, the coherent optical phonon amplitude is generated within the pump pulse duration or one oscillational period of the phonons, usually within 100 fs.
However, in this case, the coherent oscillation amplitude [Fig. \ref{reflection}(b)] increases for approximately 1 ps (almost 10 oscillations).
This behavior may be interpreted as the superposition of several dumped oscillations with slightly different frequencies and phases:
\begin{eqnarray}
F(t) &=& \sum_{i=1}^3 A_i \exp(-\frac{t}{\tau_i}) \cos (\omega_i t + \theta_i), 
\end{eqnarray}
where $t$ denotes the delay between pump and probe pulses, $A_i$ the phonon amplitude, $\omega_{i}$ the angular frequency, and $\theta_i$ the phase for the $i$-th oscillation. 

\begin{figure}[t]
    \centering
    \includegraphics[width=8cm]{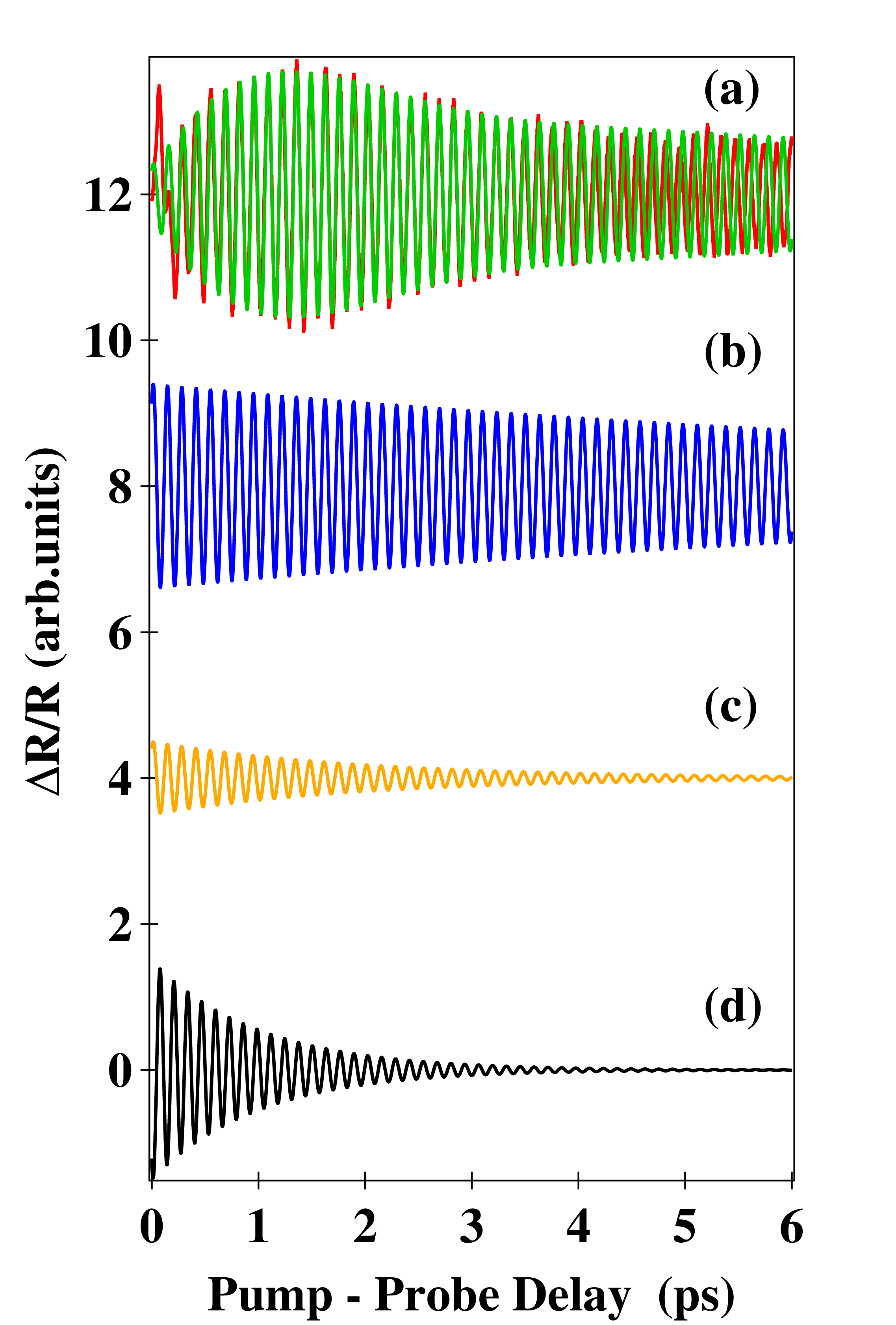}
    \vspace{1cm}
    \caption{Simulation of oscillations of coherent phonons in WSe$_2$. 
    The red curve is the experimentally obtained transient reflectivity. The blue, yellow and black curves are oscillations with frequencies 7.45, 7.49, and 7.7 THz, respectively. The green curve is the sum of the blue and red curves. }
    \label{sim}
\end{figure}
Using the parameters: 
$A_1 = 1.4$, $\omega_1 = 2 \pi \times 7.45 $ THz, $\tau_1 = 10 $ ps, $\theta_1 = 1.8 \pi$,
$A_2 = 0.5$, $\omega_2 = 2 \pi \times 7.49 $ THz, $\tau_2 = 2 $ ps, and $\theta_2 = 1.8\pi$,
$A_3 = 1.5$, $\omega_3 = 2 \pi \times 7.7 $ THz, $\tau_3 = 1 $ ps, and $\theta_3 = 0.8\pi$,
the calculated oscillation closely represents the transient reflection (Fig. \ref{sim}).
This simulation reproduces the rise of the coherent oscillation.
Notably, the phase of the 7.7-THz oscillation differs by $\pi$ from that of the 7.45 and 7.49-THz oscillations.

The Fourier spectrum [Fig. \ref{spectral}(a)] reveals low-frequency (4.00 THz) and high-frequency (11.56 THz) modes.
The low-frequency oscillation has also been reported in transient transmission experiments using tens -of-femtosecond pulses \cite{Joeng2016}. 
In our experiment, the high-frequency oscillation is detected using the \textcolor{black}{ultrashort ($\sim$20 fs) near infrared pulses}, as shorter pulses can excite the higher-frequency phonons.
Both the low-frequency and high-frequency peaks are found in the Raman spectrum and have been assigned to coupled modes \cite{Joeng2016}.

\section{Conclusions}
Our investigation of coherent phonons in bulk WSe$_2$ using  \textcolor{black}{ultrashort ($\sim$20 fs) near infrared pulses} has yielded several significant findings.
The time-resolved reflectivity data show a rise in oscillation with frequencies around 7.5 THz.
The behavior is well reproduced by a simulation superimposing three oscillations (with frequencies 7.45, 7.49 and 7.7 THz), each with different phases.
The Fourier transform spectrum reveals small peaks at 4.0 and 11.5 THz along with the intense peaks at around  7.5 THz, providing a comprehensive view of the coherent phonon dynamics in WSe$_2$.

\section*{Acknowledgments}
We thank Y. Konno, M. Hirose, and T. Nishimura for their help in the experiments.
We thank the Open Facility Center of the Institute of Science Tokyo for measuring Raman spectra.
This work was partially supported by JSPS KAKENHI (under Grants 21K18904, 21H04669, 22J23231, 22H01984, 22KJ1342, and 23K23252), Design \& Engineering by Joint Inverse Innovation for Materials Architecture; MEXT, and Collaborative Research Projects of Materials and Structures Laboratory, Institute of Integrated Research, Institute of Science Tokyo. We thank Richard Haase, PhD, from Edanz (https://jp.edanz.com/ac) for editing the English text of a draft of this manuscript.


\end{document}